\documentclass[superscriptaddress,secnumarabic,nobibnotes,aps,prd,showkeys,
showpacs,nofootinbib,preprint]{revtex4}

\usepackage{graphicx}
\usepackage{bm}
\usepackage{amsmath}
\usepackage{amsfonts}
\usepackage{amssymb}
\usepackage{epstopdf}
\usepackage{natbib}%
\newcommand{\bee}{\begin{equation}}
\newcommand{\eee}{\end{equation}}
\newcommand{\eaa}{\end{eqnarray}}
\newcommand{\baa}{\begin{eqnarray}}
\def\ni{\noindent}

\usepackage{color}

\begin{document}

\title{Tsallis and Kaniadakis statistics from a point of view of the holographic equipartition law}

\author{Everton M. C. Abreu}\email{evertonabreu@ufrrj.br}
\affiliation{Grupo de F\' isica Te\'orica e Matem\'atica F\' isica, Departamento de F\'{i}sica, Universidade Federal Rural do Rio de Janeiro, 23890-971, Serop\'edica, RJ, Brazil}
\affiliation{Departamento de F\'{i}sica, Universidade Federal de Juiz de Fora, 36036-330, Juiz de Fora, MG, Brazil}
\author{Jorge Ananias Neto}\email{jorge@fisica.ufjf.br}
\affiliation{Departamento de F\'{i}sica, Universidade Federal de Juiz de Fora, 36036-330, Juiz de Fora, MG, Brazil}
\author{Albert C. R. Mendes}\email{albert@fisica.ufjf.br}
\affiliation{Departamento de F\'{i}sica, Universidade Federal de Juiz de Fora, 36036-330, Juiz de Fora, MG, Brazil}
\author{Alexander Bonilla}\email{abonillafisica.ufjf.br}
\affiliation{Departamento de F\'{i}sica, Universidade Federal de Juiz de Fora, 36036-330, Juiz de Fora, MG, Brazil}

\pacs{51.10.+y, 05.20.-y, 98.65.Cw }
\keywords{Tsallis statistics, Kaniadakis statistics, holographic equipartition law}

\begin{abstract}
\noindent In this work,  we have illustrated the difference between both Tsallis and Kaniadakis entropies through cosmological models obtained from the formalism proposed by Padmanabhan, which is called holographic equipartition law. Similarly to the formalism proposed by Komatsu, we have obtained an extra driving constant term in the Friedmann equation if we deform the Tsallis entropy  by Kaniadakis' formalism. We have considered initially Tsallis entropy as the Black Hole (BH) area entropy. This constant term  may lead the universe to be in an accelerated or decelerated mode. On the other hand, if we start with the Kaniadakis entropy as the BH area entropy and then by modifying the Kappa expression by Tsallis' formalism, the same absolute value  but with opposite sign is obtained. In an opposite limit, no driving inflation term of the early universe was derived from both deformations.
\end{abstract}
\date{\today}

\maketitle

Tsallis' statistics \cite{tsallis}, which is an extension of Boltzmann-Gibbs's (BG) statistical theory, defines a nonextensive (NE), i.e., nonadditive entropy as
\begin{eqnarray}
\label{nes}
S_q =  k_B \, \frac{1 - \sum_{i=1}^W p_i^q}{q-1}\;\;\;\;\;\;\qquad \Big(\,\sum_{i=1}^W p_i = 1\,\Big)\,\,,
\end{eqnarray}

\ni where $p_i$ is the probability of a system to exist within a microstate, $W$ is the total number of configurations (microstates) and 
$q$, known in the current literature as the Tsallis parameter or NE  parameter, is a real parameter which measures the degree of nonextensivity. 
The definition of entropy in Tsallis statistics carries the standard properties of positivity, equiprobability, concavity and irreversibility. This approach has been successfully used in many different physical system. For instance, we can mention the Levy-type anomalous diffusion \cite{levy}, turbulence in a pure-electron plasma \cite{turb} and gravitational systems {\color{red} \cite{sys,sys2,eu}}.
It is noteworthy to affirm that Tsallis thermostatistics formalism has the BG statistics as a particular case in the limit $ q \rightarrow 1$ where the standard additivity of entropy can be recovered. 

In the microcanonical ensemble, where all the states have the same probability, Tsallis' entropy reduces to \cite{te}
\begin{eqnarray}
\label{micro}
S_q=k_B\, \frac{W^{1-q}-1}{1-q},
\end{eqnarray}
where in the limit $q \rightarrow 1$ we recover the usual Boltzmann entropy formula, $S=k_B\, \ln {W}$.

On the other hand, the well known Kaniadakis statistics  \cite{kani1}, also refereed as $\kappa$-statistics, analogously to Tsallis thermostatistics model, generalizes the usual BG statistics initially by introducing both the $\kappa$-exponential and $\kappa$-logarithm defined respectively by
\begin{eqnarray}
\label{expk}
exp_\kappa(f)=\Big( \sqrt{1+\kappa^2 f^2}+\kappa f \Big)^\frac{1}{\kappa},
\end{eqnarray}
\begin{eqnarray}
\label{logk}
\ln_\kappa(f)=\frac{f^\kappa-f^{-\kappa}}{2\kappa},
\end{eqnarray}

\ni and the following property can be satisfied, namely,
\begin{eqnarray}
\ln_\kappa\Big(exp_\kappa(f)\Big)=exp_\kappa\Big(\ln_\kappa(f)\Big)\equiv f.
\end{eqnarray}

\ni From Eqs. (\ref{expk}) and (\ref{logk}) we can notice that the $\kappa$-parameter twists the standard definitions of the exponential and logarithm functions.

The $\kappa$-entropy, connected to this $\kappa$-framework, can be written as
\begin{eqnarray}
S_\kappa=- k_B \sum_i^W  \,\frac{p_i^{1+\kappa}-p_i^{1-\kappa}}{2\kappa},
\end{eqnarray}

\ni which recovers the BG entropy in the limit $\kappa \rightarrow 0$. It is relevant to comment here that the $\kappa$-entropy satisfies the properties concerning concavity, additivity and extensivity. The $\kappa$-statistics has thrived when applied in many experimental scenarios. As an example we can cite cosmic rays  \cite{Kanisca1} and cosmic effects  \cite{aabn-1}, quark-gluon plasma  \cite{Tewe}, kinetic models describing a gas of interacting atoms and photons  \cite{Ross} and financial models  \cite{RBJ}.

Using the microcanonical ensemble definition, where all the states have the same probability, Kaniadakis' entropy reduces to \cite{kani1,Kanisca1}
\begin{eqnarray}
\label{microk}
S_\kappa=k_B\, \frac{W^\kappa-W^{-\kappa}}{2\kappa}\;,
\end{eqnarray}
where in the limit $\kappa \rightarrow 0$ we recover the usual Boltzmann entropy formula, $S=k_B\, \ln {W}$.

In order to illustrate the main difference between the Tsallis and Kaniadakis formalism we begin our formalism by considering the Tsallis microcanonical entropy formula,  
Eq.(\ref{micro}), writing as 

\begin{eqnarray}
\label{ttm}
S_t =k_B\, t \; ,
\end{eqnarray}
where $t$ is defined as

\begin{eqnarray}
\label{td}
t\equiv \frac{W^\lambda-1}{\lambda} \; ,
\end{eqnarray}
and 

\begin{equation}
\label{lamb}
\lambda\equiv 1-q \;.
\end{equation}

\ni Here it is important to mention that the signal of the $\lambda$ parameter defined in Eq.(\ref{lamb}) depends on the value of the nonextensive parameter $q$. 

Writing Kaniadakis' entropy formula, Eq.(\ref{microk}), as a function of $t$ defined in Eq.(\ref{td}), we have

\begin{eqnarray}
\label{kps}
S_\kappa= \frac{k_B}{2\lambda}\left[1+\lambda t
-\frac{1}{1+\lambda t}\right]
=\frac{k_B}{2} \left[\frac{2t+\lambda t^2}{1+\lambda t}\right].
\end{eqnarray}
We can expand (\ref{kps}) in a power series of $t$ where we obtain

\begin{eqnarray}
S_\kappa= k_B \left [t-\frac{1}{2}\lambda\, t^2 +\frac{1}{2}\lambda^2\, t^3-...\right]\;.
\end{eqnarray}

\ni Considering now the Kaniadakis microcanonical entropy formula,  
Eq.(\ref{microk}), written as 

\begin{eqnarray}
\label{ska}
S_\kappa= k_B \, \kappa \;,
\end{eqnarray}
where $\kappa$ is defined as

\begin{eqnarray}
\label{kf}
\kappa\equiv\frac{W^\lambda-W^{-\lambda}}{2\lambda}\;,
\end{eqnarray}
and writing Tsallis' entropy formula, Eq.(\ref{micro}), as a function of $\kappa$ defined in Eq.(\ref{kf}), we have that

\begin{eqnarray}
\label{tps}
S_t=k_B\left[\kappa+\frac{\left( 1+\lambda^2 \kappa^2 \right)^{\frac{1}{2}}}
{\lambda}-\frac{1}{\lambda}\right],
\end{eqnarray}

\ni where the plus sign before the square root guarantees us that when $\lambda=0$ the entropy, Eq.(\ref{tps}), is not deformed.
We can expand (\ref{tps}) in a power series of $\kappa$ where we obtain

\begin{eqnarray}
S_t=k_B\left[\kappa+\frac{1}{2}\lambda\, \kappa^2-\frac{1}{8} \lambda^3\, \kappa^4+...\right]\,.
\end{eqnarray}

Following the same line of Komatsu \cite{komatsu}, the holographic equipartition law proposed by Padmanabhan \cite{Pad1} considers
the time rate of change of the cosmic volume as

\begin{eqnarray}
\label{heq}
\frac{dV}{dt}=L_p^2 c\, (N_{sur}-\epsilon N_{bulk}),
\end{eqnarray}
where $L_p$ is the Planck length, $c$ is the velocity of light, $N_{sur}$ is the
number of degrees of freedom on the spherical surface of Hubble radius $r_H$ and
$N_{bulk}$ is the number of degrees of freedom in the bulk. {\color{red}The Hubble horizon radius is given by 

\begin{eqnarray}
\label{rh}
r_H=\frac{c}{H},
\end{eqnarray}
where $H$ is the Hubble parameter defined as $H\equiv \frac{da/dt}{a(t)}=\frac{\dot{a}(t)}{a(t)}$, and $a(t)$ is the scale factor of the Friedmann-Lema\^itre-Robertson-Walker metric.
Consequently the Hubble volume $V$ can be written as}

\begin{eqnarray}
\label{vol}
V=\frac{4\pi}{3} \, r_H^3 = \frac{4\pi}{3} \left(\frac{c}{H}\right)^3.
\end{eqnarray}
The equipartition theorem determines the number of degrees of freedom in the
bulk which can be written as

\begin{eqnarray}
\label{nbulk}
N_{bulk}=\frac{\color{red}|E|}{\frac{1}{2} k_B T},
\end{eqnarray}
where $\color{red}|E|$ is the {\color{red}nonnegative} value of {\color{red} the} Komar energy inside the Hubble volume $V$
which is given by

\begin{eqnarray}
\label{energy}
{\color{red}{|E|}}=\epsilon (\rho c^2+3p) V,
\end{eqnarray}

\ni {\color{red} where $\rho$ and $p$ are the energy density and the pressure respectively.} The accelerated universe (dark energy) corresponding to $(\rho c^2+3p) < 0$ which 
implies that $\epsilon=+1$ in {\color{red}(\ref{energy}). On the other hand, $(\rho c^2+3p) > 0$ corresponds to a decelerating universe (matter and radiation universe) and consequently we have $\epsilon=-1$. For more details see reference \cite{komatsu}}. The temperature T on the horizon is written as

\begin{eqnarray}
\label{temp}
T=\frac{\hbar H}{2\pi k_B}.
\end{eqnarray}

\ni The number of degrees of freedom on the spherical surface is given by

\begin{eqnarray}
\label{nsur}
N_{sur}=\frac{4S_H}{k_B},
\end{eqnarray}

\ni where $S_H$ is the entropy on the Hubble horizon. When the entropy on the Hubble horizon, $S_H$, is equal to the Bekenstein-Hawking entropy, $S_{BH}$, i.e.
\begin{equation}
S_H=S_{BH}=\frac{k_B A_H}{4L_p^2}\;,
\end{equation}
where $A_H$ is the surface area of the sphere with the Hubble horizon(radius) $r_H$ defined in Eq.(\ref{rh}) and $L_p$ is the Planck length,
then  $N_{sur}$ is the usual number of degrees of freedom on the surface that is $N_{sur}=\frac{A}{L_p^2}$. The Bekenstein-Hawking entropy, $S_{BH}$, can be written as

\begin{eqnarray}
\label{sbh}
S_{BH}=\frac{k_B A_H}{4\color{red}L_p^2}=\frac{\pi k_B c^2}{L_p^2 H^2}=\frac{k'}{H^2},
\end{eqnarray}
where $A_H=4\pi r_H^2$ and  $k'\equiv \frac{\pi k_b c^2}{L_p^2}$.

Here we would like to mention that it is possible to consider the microstates number  as $$ W=b\, \Big(\frac{A_H}{4\color{red}L_p^2}\Big)^\alpha\,\,,$$  where
$b$ and $\alpha$ are, at first, undetermined parameters, in both Tsallis and Kaniadakis entropies, Eqs. (\ref{micro}) and (\ref{microk}) respectively.
Adjusting conveniently the constants $b$ and $\alpha$, then the BH entropy, Eq.(\ref{sbh}), can be retrieved from  Tsallis and Kaniadakis statistics. For more details see reference \cite{epl}.

In order to derive the Friedmann equation by the holographic equipartition law,
we will start by calculating the left-hand side of Eq. (\ref{heq}). Substituting Eq. (\ref{vol}) into (\ref{heq}) we obtain

\begin{eqnarray}
\label{dvt}
\frac{dV}{dt}=-4\pi c^3 \left(\frac{\dot{H}}{H^4}\right).
\end{eqnarray}

\ni Then, using Eqs.(\ref{nbulk}), (\ref{energy}), (\ref{temp}), (\ref{nsur}) and (\ref{dvt}) into Eq. (\ref{heq}) we have

\begin{eqnarray}
\label{frideq}
\frac{\ddot{a}}{a}
=-\frac{4\pi G}{3} \left(\rho+\frac{3p}{c^2}\right)+ H^2\left(1-
\frac{S_H }{S_{BH}}\right),
\end{eqnarray}

\ni where we have used
\begin{eqnarray}
\label{aa}
\frac{\ddot{a}}{a}= \dot{H}+H^2,
\end{eqnarray}

\ni and $\epsilon=+1$ which corresponds to an accelerated universe.

From Eq.(\ref{frideq}) we can observe that the driven term that indicates an accelerated universe is given by
\begin{eqnarray}
f(H)=H^2 \bigglb( 1-\frac{S_H}{S_{BH}} \biggrb) \,\,.
\end{eqnarray}

\ni Considering $S_H=S_\kappa$,  Eq.(\ref{kps}), $S_{BH}=S_t$, Eq.\eqref{ttm}, and using (\ref{sbh})  we have
\begin{eqnarray}
\label{fh}
f(H)=\frac{k' \lambda}{2k_B(1+\lambda t)}.
\end{eqnarray}

\ni Expanding Eq.(\ref{fh}) in a power series of $\lambda$ and taking only
the linear term in $\lambda$ we obtain

\begin{eqnarray}
\label{ack}
f(H)=\frac{k' \lambda}{2k_B(1+\lambda t)}= \frac{k' \lambda}{2k_B}
 (1+\lambda t)^{-1} \approx \frac{k' \lambda}{2 k_B}.
\end{eqnarray}

\ni Hence, we can mention that when we deform Tsallis by Kaniadakis entropy, Eq.(\ref{kps}), a cosmological type constant term appears, Eq.(\ref{ack}). 
Depending on the $\lambda$ sign, this constant may be positive or negative. This result illustrates a difference between Tsallis and Kaniadakis formalism. Moreover, this result is the same obtained by Komatsu\cite{komatsu} 
when R\'enyi entropy is deformed by Tsallis expression.


Considering now $S_H=S_t$,  Eq.(\ref{tps}), $S_{BH}=S_\kappa $, Eq.(\ref{ska}), and using (\ref{sbh}) we have

\begin{eqnarray}
\label{fhk}
f(H)=\frac{k'}{k_B} \, \frac{1-(1+\lambda^2 \kappa^2)^\frac{1}{2}}{\lambda \, \kappa^2}.
\end{eqnarray}
Expanding Eq.(\ref{fhk}) in a power series $\lambda$ and taking only
the linear term in $\lambda$ we obtain

\begin{eqnarray}
\label{ackk}
f(H)=\frac{k'}{k_B} \, \frac{1-(1+\lambda^2\, \kappa^2)^\frac{1}{2}}{\lambda \, \kappa^2} \approx - \frac{k' \lambda}{2 k_B}.
\end{eqnarray}

\ni So, when we deform the Kaniadakis entropy by the Tsallis one, Eq.(\ref{tps}),
a cosmological constant term also appears, Eq.(\ref{ackk}), which is the same absolute value obtained in (\ref{ack}) but with opposite sign. This fact can again illustrate a difference between Kaniadakis and Tsallis entropy.

We can analyze Eq.(\ref{fh}) in the limit $\lambda t >> 1$ where we obtain

\begin{eqnarray}
\label{lim2}
f(H)=\frac{k' \lambda}{2k_B(1+\lambda t)}=\frac{k'}{2 k_b t (1+\frac{1}{\lambda t})}=\frac{k'}{2 k_B t} (1+\frac{1}{\lambda t})^{-1}\approx \frac{k'}{2 k_B t}\approx \frac{H^2}{2}.
\end{eqnarray}

\ni From Eq.(\ref{lim2}) we can observe that a $H^2$ like term is obtained. This result is similar to the one obtained by Komatsu\cite{komatsu} and we can mention that this driving term does not produce inflation of the early universe  where a $H^4$ like term should be required.

From the driving term, Eq.(\ref{fhk}), we can also perform the limit $\lambda \kappa >> 1$ and the result is
\begin{eqnarray}
\label{lim3}
f(H)=\frac{k'}{k_B} \, \frac{1-(1+\lambda^2 \kappa^2)^\frac{1}{2}}{\lambda \kappa^2}=\frac{k'}{k_B} \frac{1-\lambda \kappa(1+\frac{1}{\lambda^2 \kappa^2})^\frac{1}{2}}{\lambda \kappa^2}\approx - \frac{k'}{k_B \kappa}\approx - H^2.
\end{eqnarray}

\ni From Eq.(\ref{lim3}) we can observe that in the limit $\lambda \kappa >> 1$, we have $f(H)\approx - H^2$. Then, the term obtained when we consider Kaniadakis' entropy deformed by Tsallis' approach  does not also produce inflation of the early universe  where a $H^4$ like term should be needed.

To conclude, in this paper we have studied differences between the Tsallis and Kaniadakis entropy from the point of view of the holographic equipartition formalism. Initially, in the limits $\lambda t << 1$ and $\lambda \kappa << 1$ , the Tsallis and Kaniadakis entropies both differ by a cosmological term represented by Eqs.(\ref{ack}) and (\ref{ackk}). Then, when we deform the Tsallis entropy by the Kaniadakis approach, in the limit $\lambda t >> 1$, the resulting difference term, Eq.(\ref{lim2}), does not correspond to a inflation term of the early universe. In an inverse procedure, when we deform the Kaniadakis entropy by the Tsallis approach, in the limit $\lambda \kappa >> 1$, also no inflation term in a early universe, that is Eq.(\ref{lim3}), is obtained.

\section{Acknowledgments}

\ni The authors thank CNPq (Conselho Nacional de Desenvolvimento Cient\' ifico e Tecnol\'ogico), Brazilian scientific support federal agency, for partial financial support, Grants numbers 302155/2015-5 (E.M.C.A.) and 303140/2017-8 (J.A.N.). E.M.C.A. thanks the hospitality of Theoretical Physics Department at Federal University of Rio de Janeiro (UFRJ), where part of this work was carried out.

\end{document}